\documentclass[prl,twocolumn,superscriptaddress,reprint, amsmath,amssymb]{revtex4-1}

\usepackage{graphicx}
\usepackage{dcolumn}
\usepackage{bm}

\usepackage{tikz}

\newcommand{\beq}{\begin{equation}}
\newcommand{\eeq}{\end{equation}}

\newcommand{\eps}{\varepsilon}

\renewcommand{\rho}{\varrho}
\renewcommand{\theta}{\vartheta}
\renewcommand{\phi}{\varphi}

\renewcommand{\Re}{{\rm Re} \,}
\renewcommand{\Im}{{\rm Im} \,}

\newcommand{\wegdamit}[1]{} 

\newlength{\lwveryfine}   
\newlength{\lwfine}   
\newlength{\lwnormal} 
\newlength{\lwthick}  
\newlength{\lwverythick}  

\begin{document}

\preprint{APS/123-QED}

\title{Alternating Phase Focusing for Dielectric Laser Acceleration}

\author{Uwe Niedermayer}
 \email{niedermayer@temf.tu-darmstadt.de}
\author{Thilo Egenolf}
\author{Oliver Boine-Frankenheim}
\affiliation{%
Technische Universit\"at Darmstadt, Schlossgartenstrasse 8, D-64289 Darmstadt, Germany
}%


\author{Peter Hommelhoff}
 \affiliation{Department Physik, Friedrich-Alexander-Universit\"at Erlangen-N\"urnberg (FAU), Staudtstrasse 1, D-91058 Erlangen
}

\date{\today}

\begin{abstract}
The concept of Dielectric Laser Acceleration (DLA) provides highest gradients among non-plasma particle accelerators. However, stable beam transport and staging have not been shown experimentally yet. 
We present a scheme that confines the beam longitudinally and in one transverse direction. Confinement in the other  direction is obtained by a single conventional quadrupole magnet. 
Within the small aperture of 420 nm we find the matched distributions, which allow an optimized injection into pure transport, bunching, and accelerating structures. 
The combination of these resembles the photonics analogue
of the Radio Frequency Quadrupole (RFQ), but since our setup is entirely two-dimensional, it can be manufactured on a microchip by lithographic techniques. This is a crucial step towards relativistic electrons in the MeV range from low-cost, handheld devices.
\end{abstract}

\pacs{Valid PACS appear here}
\maketitle


Since Dielectric Laser Acceleration (DLA) of electrons has been proposed in 1962~\cite{Shimoda1962ProposalMaser,Lohmann1962ElectronWaves}, the development of photonic nano-structures and the control of ultrashort laser pulses has advanced significantly (see~\cite{England2014DielectricAccelerators} for an overview). Phase synchronous acceleration was experimentally demonstrated first in 2013~\cite{Peralta2013DemonstrationMicrostructure., Breuer2013Laser-BasedStructure}. 
Record gradients, more than an order of magnitude higher than in conventional accelerators, were achieved meanwhile both for relativistic~\cite{Wootton2016DemonstrationPulses} and low-energy electrons~\cite{Leedle2015LaserStructure}.   
These gradients, so far, express themselves only in the generation of energy spread, not as a coherent acceleration. 
Moreover, the interaction length is limited to the Rayleigh length, after which the electron beam defocuses and hits the small (sub-micrometer) aperture. During synchronous acceleration, there are additional defocusing forces which cannot be overcome by magnetic focusing only~\cite{Ody2017FlatAccelerators}.

In this letter we show a laser-based scheme which allows transport and acceleration of electrons in dielectric nano-structures over arbitrary lengths. It is applicable to changing DLA period lengths, which is required to accelerate subrelativistic electrons. 
Moreover, we find the maximum tolerable emittances and beam envelopes in DLA beam channels. Another advancement of our scheme is ballistic bunching of subrelativistic electrons down to attosecond duration, while the beam remains transversely confined.
This paves the way to a low-cost accelerator on a microchip, providing MeV electrons from a small-scale, potentially handheld device. 
 
Our scheme uses only one spatial harmonic, namely the synchronous one, but its magnitude and phase change along the DLA grating. This is interpreted as a time dependent focusing potential. 
A focusing concept using non-synchronous spatial harmonics of travelling waves was presented by Naranjo et al.~\cite{Naranjo2012StableHarmonics}. They derived
stability due to retracting ponderomotive forces from the non-synchronous spatial harmonics, while the synchronous one serves for acceleration.
Our description is in the co-moving real space, as compared to Naranjo's description in the spatial frequency domain. This supports changes of all grating-related quantities, while the Courant-Snyder (CS) theory~\cite{Courant1958TheorySynchrotron} from conventional accelerator physics is still applicable. 
Stable beam confinement is achieved by Alternating Phase Focusing (APF), which was already developed in the 1950s for ion acceleration~\cite{Fainberg1956AlternatingFocusing}. However, the later developed Radio Frequency Quadrupole (RFQ) cavities turned out to have better performance, especially at high current beams. Thus, APF was rejected in favor of the RFQ and only rarely implemented~\cite{Wangler2008RFAccelerators}. Since 3D structures as RFQs are not feasible for lithographic fabrication on a microchip, we recover APF in this letter in order to stabilize DLA.

\begin{figure}[h]
\centering
\includegraphics[width=0.45\textwidth]{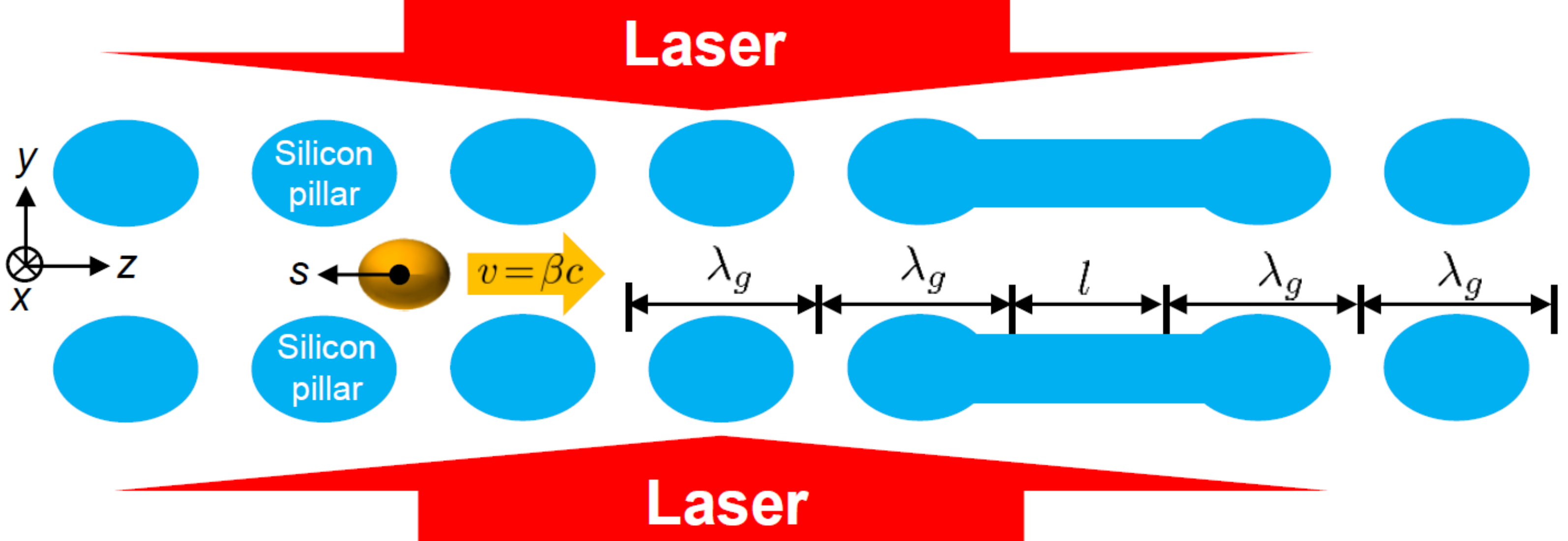}
\includegraphics[width=0.48\textwidth]{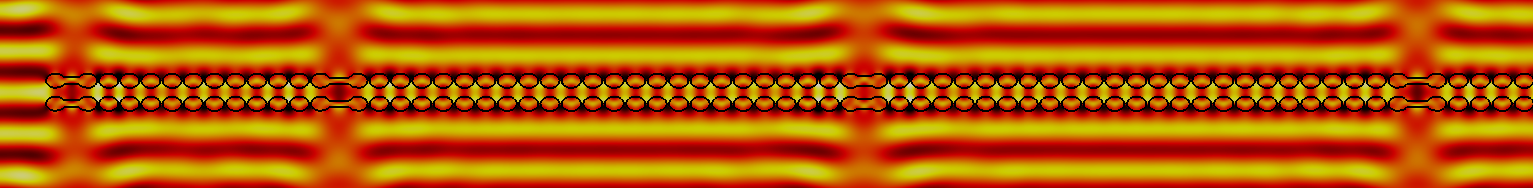}
\caption{Schematic view of a dual pillar DLA structure and particle bunch around a reference particle (top) and simulation of the longitudinal time harmonic electric field (bottom).}
\label{fig:Overview}
\end{figure}
We use standing wave dual pillar structures~\cite{Leedle2015LaserStructure} as shown in Fig.~\ref{fig:Overview}, but our scheme can also be applied to Bragg cavity structures~\cite{Niedermayer2017DesigningChip}.
The z-polarized lasers, incident from both lateral sides, are modeled as  plane waves with wavelength $\lambda_0=2\,\mu$m. In practice, they can be realized as pulse-front-tilted profiles~\cite{Cesar2018OpticalAccelerator,Cesar2018EnhancedLaser,Kozak2018UltrafastNanostructures}, extending the interaction length compared to  non-tilted pulses.
The Hamiltonian for single particle motion in the DLA is~\cite{Niedermayer2017BeamScheme}
\begin{equation}
H=\frac{1}{2m_e \gamma} (p_x^2+p_y^2+(\Delta p_z/\gamma)^2)+V,
\label{Hamiltonian}
\end{equation}
where $\gamma=(1-\beta^2)^{-1/2}$ is the reference mass factor, $m_e$ the electron mass, $p_x$, $p_y$ the transverse momenta, and $\Delta p_z$ the deviation of longitudinal momentum from the reference particle at fixed laser phase (black dot in Fig.~\ref{fig:Overview}). In~\cite{Niedermayer2017BeamScheme} we have shown by means of the Panofsky-Wenzel theorem~\cite{Panofsky1956SomeFields} that the time dependent potential can be written as
\begin{equation}
V=q \Im\{e_1 [\frac{\lambda_g}{2\pi}\cosh\left(\frac{\omega y}{\beta\gamma c}\right) e^{2\pi i s/\lambda_g}- is e^{i\phi_s}]\},
\label{Potential}
\end{equation}
where $\omega=2\pi c/\lambda_0$ is the laser angular frequency, $q$ is the (negative) electron charge, and $s$ is the distance of the particle behind the reference particle. 
The field strength of the resonant harmonic with the Wideroe condition $\lambda_g=\beta\lambda_0$ is $e_1$, i.e., with no loss of generality we work with the first (usually the strongest) spatial harmonic.  
The parameters $e_1$, $\phi_s$, $\beta, \gamma$ and $\lambda_g$ are allowed to vary with the time-like cell index $n$.
The synchronous phase $\phi_s$ determines the energy gain of the reference particle as function of the cell number (the acceleration ramp) as
\begin{equation}
W_{kin}(N)=W_{kin,0}+q\sum_{n=1}^N \lambda_g^{(n)} \Re\{e_1^{(n)} e^{i\phi_s^{(n)}} \},
\label{Eq:Ramp}
\end{equation}
where $W_{kin,0}=83$ keV is the injection energy. 
The cell lengths increase according to the Wideroe condition as
\begin{equation}
\frac{\lambda_g^{(n+1)}-\lambda_g^{(n)}}{\lambda_0}=\beta^{(n+1)}-\beta^{(n)}=\frac{q \lambda_0 \Re\{e_1^{(n)} e^{i\phi_s^{(n)}}\}}{m_e c^2 \gamma^{{(n)}^3}}.  
\nonumber
\end{equation} 
For a given structure the synchronous phase is thus determined as $\phi_s^{(n)}=\phi_0-\arg(e_1)^{(n)}$, where
\begin{equation}
\phi_0=\arccos\left[\frac{m_e c^2 }{q\lambda_0}\frac{\gamma^3}{|e_1|}\frac{\Delta \lambda_g}{\lambda_0}\right].
\end{equation}
In this letter we use optimized structures which provide $\phi_0$ independent of $n$ at arbitrary chirp parameter $\Delta\lambda_g$, such that the synchronous phase $\phi_s$ can be switched by a particular drift from one grating segment to another. Tying the phase $\arg(e_1)^{(n)}$ to $\lambda_g^{(n)}$ does not avoid a small drift in the normalized amplitude $|e_1^{(n)}/E_L|\approx 0.33...0.37$ (see~\cite{Niedermayer2018SupportingMaterial}), which is taken into account in the ramp (Eq.~\ref{Eq:Ramp}).

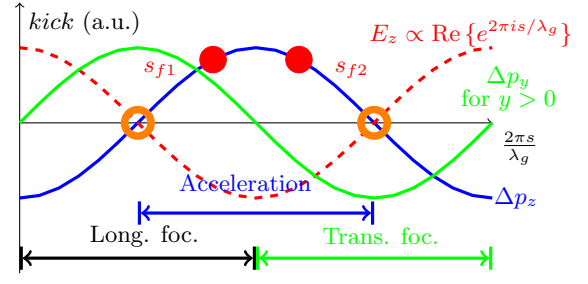
\begin{figure}[t]
\begin{tikzpicture}
  \draw[->] (0,0) -- (2*pi,0) node[below right] {$\frac{2\pi s}{\lambda_g}$};
  \draw[->] (0,-2) -- (0,1.6) node[below right] {$kick$ (a.u.)};
  \draw [blue, very thick, domain=0:2*pi] plot ({\x}, {-cos(180/pi*\x)}) ;
  \draw [red,dashed, very thick, domain=0:2*pi] plot ({\x}, {cos(180/pi*\x)}) ;
  \node [red] at (6.0,1.2) {$E_z\propto\Re \{e^{2\pi i s/\lambda_g} \} $};
  \draw [green, very thick, domain=0:2*pi] plot ({\x}, {sin(180/pi*\x)}) ;
  \node [green] at (6.5,0.6) {$\Delta p_y$};
  \node [green] at (6.5,0.3) {for $y>0$};
  \node [blue] at (6.6,-1.0) {$\Delta p_z$};
  \node [circle, draw=orange, inner sep=3pt, line width=1mm ,minimum size=10pt] (b) at (pi/2,0) {};
  \node [circle, draw=orange, inner sep=3pt, line width=1mm, minimum size=10pt] (b) at (3*pi/2,0) {};
  
  \node [circle, draw=red, fill=red, inner sep=3pt,minimum size=10pt] (b) at (pi/2+1,0.8415) {};
  \node [circle, draw=red, fill=red, inner sep=3pt,minimum size=10pt] (b) at (3*pi/2-1,0.8415) {};
    \node [red] at (pi/2+0.3,0.7) {$s_{f1}$};
  \node [red] at (3/2*pi-0.3,0.7) {$s_{f2}$};

\draw[|<->|,black, very thick] (0,-1.8) -- (pi,-1.8);
  	\draw[|<->|,green, very thick] (pi,-1.8) -- (2*pi,-1.8);
	\node [black] at (4.8-pi,-1.5) {Long. foc.};
	\node [green] at (4.8,-1.5) {Trans. foc.};
   	\draw[|<->|,blue,very thick] (1/2*pi,-1.2) -- (3/2*pi,-1.2);
  	\node [blue] at (3.0,-0.8) {Acceleration};    
\end{tikzpicture}
\caption{Overview of electron acceleration and focusing properties as function of phase. The circles denote the fixed points for different $\phi_s$.
}
\label{PhaseRelation}
\end{figure}

Earnshaw's theorem dictates that constant focusing cannot be achieved in all 3 spatial directions simultaneously~\cite{Earnshaw1842OnEther}. Thus, at least two focusing directions have to be alternating. In conventional Alvarez linacs or in synchrotrons, constant focusing is applied in the longitudinal direction and alternating quadrupole lattices provide transverse confinement~\cite{Lee2004AcceleratorPhysics}. Here we apply the alternation to the disjoint focusing phase ranges of the longitudinal plane and the non-invariant transverse plane ($y$). Jumping the reference particle by means of a fractional cell drift between the orange circles in Fig.~\ref{PhaseRelation} provides stable transport at constant energy, between the red dots we additionally obtain acceleration.
The strong acceleration defocusing in $y$ is compensated by acceleration focusing at the longitudinally unstable phase. In the invariant $x$ direction a single conventional quadrupole magnet suffices to confine the beam to the structure height~\cite{Niedermayer2018SupportingMaterial}. 

We find the fixed points of the motion by setting $\nabla V=0$ as $s_{f1}=\phi_s\lambda_g/2\pi$ and $s_{f2}=-\lambda_g/2\pi (\phi_s+2\arg(e_1))$ and define $\Delta s_1=s-s_{f1}$ and $\Delta s_2=s-s_{f2} $. Note that in the longitudinal plane for $\arg(e_1)=0$ the fixed point $s_{f1}$ is elliptic and $s_{f2}$ is hyperbolic, and vice versa in the transverse plane. Expanding $V$ to second order and omitting constant terms provides
\begin{align}
&V(x,y,s=s_{f1}+\Delta s)=-V(x,y,s=s_{f2}+\Delta s)  \nonumber \\
&=\frac{q|e_1|\lambda_g}{2\pi}
\left[\frac12 \left(\frac{\omega y}{\beta\gamma c}\right)^2
-\frac12 \left(\frac{2\pi}{\lambda_g}\Delta s\right)^2 \right] \sin(\phi_0), 
\label{SquareWell}
\end{align}
i.e., switching between $s_{f1}$ and $s_{f2}$ with $\Delta s= \Delta s_1 = \Delta s_2$ flips the sign of the potential.
Only the non-accelerating case ($\phi_0=\pi/2$) provides two interchangeable buckets, whereas a $\pi$-shifted version of the accelerating bucket will be decelerating and unstable due to mismatch with the ramp.
Hill's equations of the linearized motion are found from Eqs.~\ref{Hamiltonian} and~\ref{SquareWell} as
\begin{subequations}
\begin{align}
y''+  K y&=0 \label{Hill_y}\\
\Delta s''- K \Delta s&=0,
\end{align}
\label{EqOfMotion}
\end{subequations}
where $K=|q\omega e_1/(m_e\beta^3\gamma^3c^3 )|\sin(\phi_s)$. Note that linearization leads to decoupling of the nonlinear equations of motion, which are coupled due to Eq.~\ref{Potential}. 
The segments between two phase shifts are enumerated by $P$, such that  
\begin{equation}
\arg(e_1)(P)=
\begin{cases}
0,  P\; \mathrm{odd} \\
2\phi_0, P\; \mathrm{even}
\end{cases}
\end{equation}
leads to a sign alternation in the focusing function $K$ in Eqs.~\ref{EqOfMotion}. 
In order to switch between the two fixed points we take short drift sections denoted by $l$ and model the lattice as thick lenses of length $L^\mathrm{f}$ and $L^\mathrm{d}$. 
Each lattice cell consists of two segments and has $p$ transverse focusing and $p$ transverse defocusing elements, thus its length is given by $L=L^\mathrm{f}+l^\mathrm{f}+L^\mathrm{d}+l^\mathrm{d}$, where
\begin{subequations}
\begin{align}
L^\mathrm{f}=\sum _{n=1}^{p} \lambda_g^{(n)} &, \;\; L^\mathrm{d}=\sum _{n=p+1}^{2p} \lambda_g^{(n)}, \\
l^\mathrm{f}=(2\pi-\phi_s^{(p)})\lambda_g^{(p)}/\pi&, \;\; 
l^\mathrm{d}=(\pi-\phi_s^{(2p)})\lambda_g^{(2p)}/\pi.
\end{align}
\end{subequations}
The solution to Eqs.~\ref{EqOfMotion} is found by applying the CS formalism~\cite{Courant1958TheorySynchrotron} to the channel of thick focusing (F) and defocusing (D) elements. 
We start with a non-accelerating transport structure, i.e. $\phi_0=\pi/2$, where the lattice cells are strictly periodic. In a long lattice cell ($p\gg 1$) we can neglect the drift sections and represent it as~\cite{Niedermayer2018SupportingMaterial}
\begin{equation}
\mathbf{M}(z,L)=
\begin{cases}
\mathbf{M}_\mathrm{f}(z), &0<z<L/2 \\
\mathbf{M}_\mathrm{d}(z-L/2)\mathbf{M}_\mathrm{f}(L/2), &L/2<z<L
\nonumber
\end{cases}
\end{equation}
\begin{figure}[b]
\centering
\includegraphics[width=0.45\textwidth]{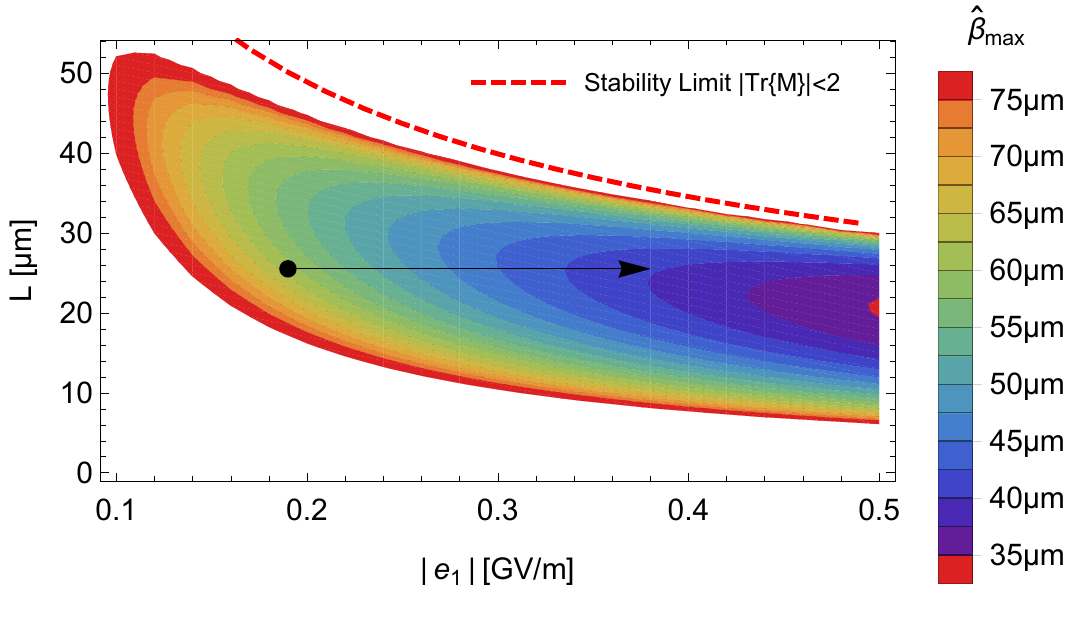}
\includegraphics[width=0.45\textwidth]{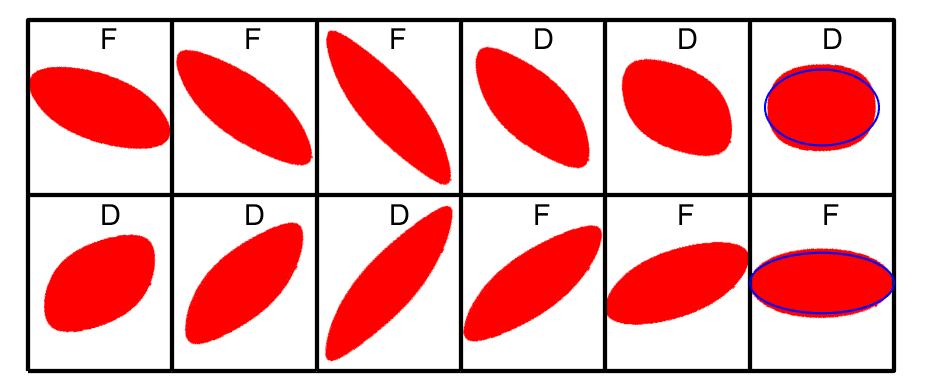}
\caption{Contours of $\hat\beta_\mathrm{max}=\hat\beta(L/4)$ in the $(|e_1|,L)$-plane (top). 
The arrow indicates the laser amplitude dependent tuning range. The transverse phase space evolution (parameters  at the black dot on top) of particles, not hitting the aperture ($\pm 0.21\,\mu$m) within 1200 DLA cells, is shown as every 2 DLA cells (bottom). The blue ellipses are the linear theory, at minimum and maximum beam size.   
}
\label{betafunctioncontour}
\end{figure}
with the length $L=(2p+1)\lambda_g$.
The phase advance per cell $\sigma$ is given for a strictly periodic FD-cell by
\begin{equation}
\cos(\sigma)=\frac12\mathrm{Tr}\{\mathbf{M}(L,L)\} =
\cos \left(\frac{\sqrt{K} L}{2}\right) \cosh \left(\frac{\sqrt{K} L}{2}\right). \nonumber
\end{equation}
The CS parameters $\eta=(\hat{\beta},\hat{\alpha},\hat{\gamma})^\mathrm{T}$ are mapped from one point to another by the matrix $\mathbf T$ (see~\cite{Niedermayer2018SupportingMaterial}) and fulfill the eigenvector relation $\eta_e=\mathbf{T}\eta_e$ for their initial values. 
For small $\sigma$, the constant $\hat\beta$ function in the smooth approximation is found from $\langle\hat\beta \rangle=L/\sigma$. However,
the most critical issue in DLA is to 
match a given emittance into the tiny aperture. Thus, the maximum of the $\hat\beta$~function, which appears at $L/4$, needs to be minimal (see Fig.~\ref{betafunctioncontour}).
The only variable parameter in an experimental setup is the laser field strength. Its tuning range from maximal admissible beam size to the structure damage threshold~\cite{Soong2012LaserAccelerators} is indicated by the black arrow. The evolution of the transverse phase space is shown below, where the particles were initially arranged on a cartesian grid and only the long term surviving ones are displayed in red. For simplicity, this simulation starts at $L/4$, 
in order to avoid correlations in the conjugate variables.
This plot uses zero bunch length, but stability is also attained for an unbunched beam, see the video in~\cite{Niedermayer2018SupportingMaterial}.
The blue ellipses indicate the strictly periodic linear case, which is slightly smaller in area, due the cosh-potential in Eq.~\ref{Potential} being steeper than the square well in Eq.~\ref{SquareWell}.
In the linear case, the single particle emittances are invariants  
\begin{subequations}
\begin{align}
\eps(y,y')&=\hat \gamma y^2 +2\hat\alpha yy' + \hat\beta y'^2, \\
\eps_L(\Delta s,\Delta s')&=\hat \gamma_L \Delta s^2 +2\hat\alpha_L \Delta s\Delta s' + \hat\beta_L \Delta s'^2,
\end{align}
\end{subequations}
where $\Delta s'=\Delta W/(m_e\gamma^3\beta^2c^2)$, and we introduce longitudinal CS-functions as a half lattice cell shift of the transverse ones, $\eta_L(z)=\eta(z-L/2)$.

\begin{figure}[t]
\centering
\includegraphics[width=0.5\textwidth]{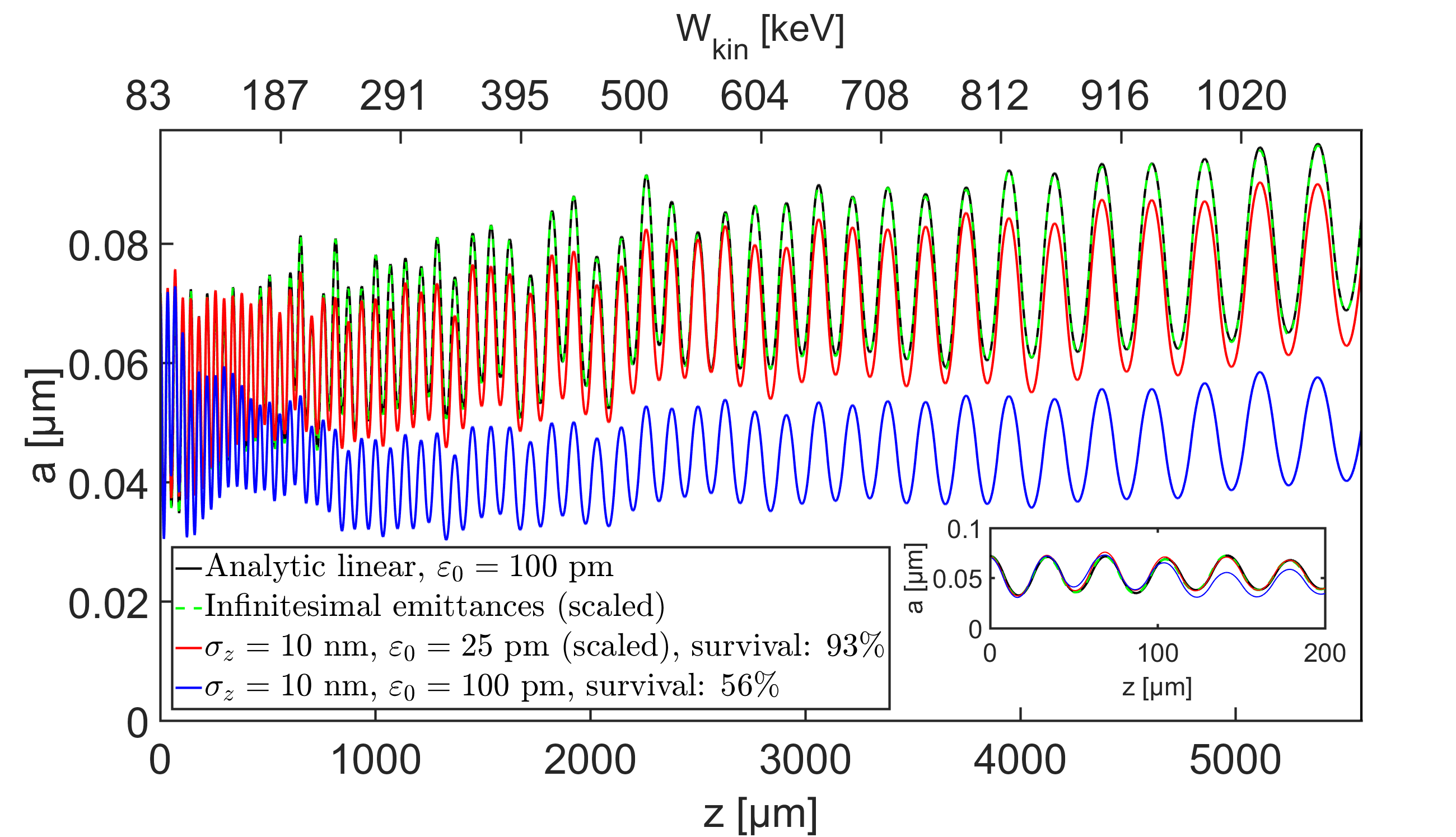}
\caption{Analytical (Eq.~\ref{Eq:Envelope}) and numerical (rms) beam envelopes, scaled to identical initial beam size at $\eps=100$ pm. 
The inset is a zoom-in of the beginning.}
\label{envelope}
\end{figure}

An accelerating lattice can be attained by taking the initial values from the eigenvalue solution and successively multiplying the segment maps as
$
\eta_N=
\mathbf{T}_N...\mathbf{T}_1
\eta_e
$
to it. In non-periodic lattices the longitudinal CS-functions have to be calculated individually with the same procedure.
\begin{figure}[b]
\centering
\includegraphics[width=0.5\textwidth]{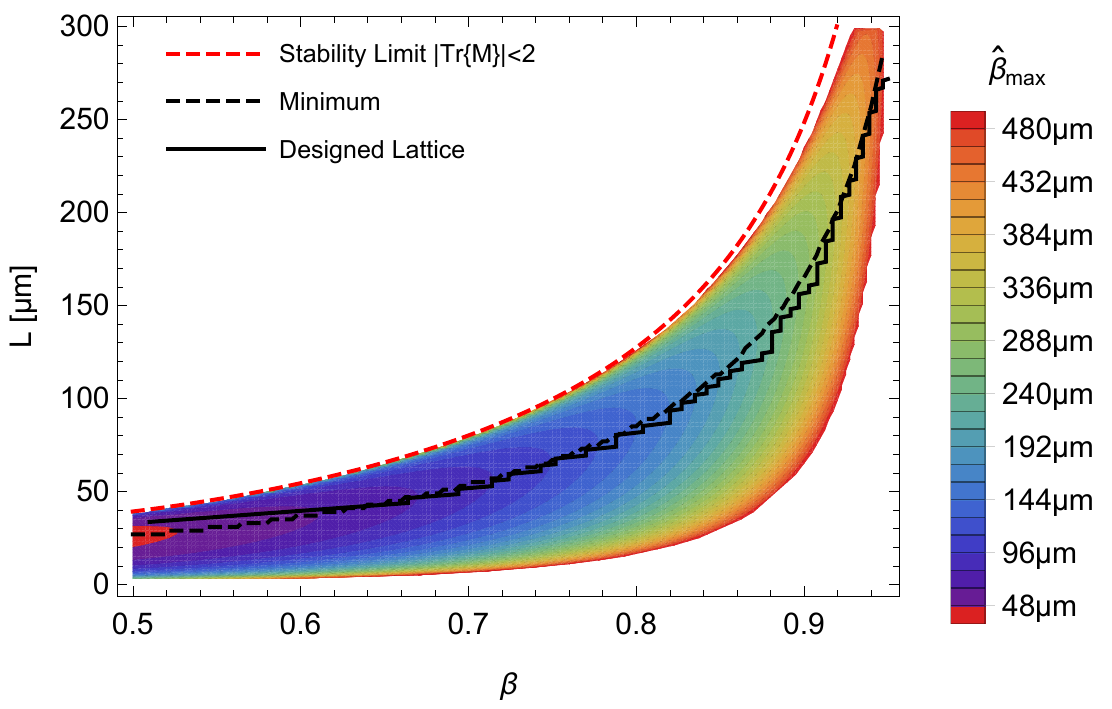}
\caption{Contours of $\hat\beta_\mathrm{max}=\hat\beta(L/4)$ in the $(\beta,L)$-plane. The designed accelerator lattice is a trade-off between following the minimum and minimizing the mismatch at the jumps.}
\label{betarelopt}
\end{figure}
If the change in length from one period to another is small, the $\hat\beta$ function can be approximated by the eigenvalue solution in each cell, which is, however, discontinuous at the boundaries. 
The line of minimal maximum of the $\hat\beta$ function in Fig.~\ref{betarelopt} is followed only approximately. On the other hand, there is adiabatic emittance damping due to momentum conservation. All together the beam envelope can be written as~\cite{Lee2004AcceleratorPhysics}
\begin{equation}
a(z)=\sqrt{\hat\beta(z)\frac{\eps_0\beta_0\gamma_0}{\beta(z)\gamma(z)}},
\label{Eq:Envelope}
\end{equation}
where the 0-indices denote initial values. Acceleration from 83~keV to 1~MeV at $\phi_0=4\pi/3$ with an average gradient of 187~MeV/m and 500~MV/m incident laser field strength from both sides is shown to be well confined within the physical aperture of $\pm 0.21\,\mu$m in Fig.~\ref{envelope}.   The analytical and numerical results coincide for infinitesimally low emittance. At small but achievable emittances~\cite{Ehberger2015HighlyTip,Feist2017UltrafastBeam}, we obtain $56\%$ transmission for $\eps_{0}=100$~pm (see video~\cite{Niedermayer2018SupportingMaterial}) and $93\%$ for $\eps_{0}=25$~pm.
\begin{figure}[b]
\centering
\includegraphics[width=0.225\textwidth]{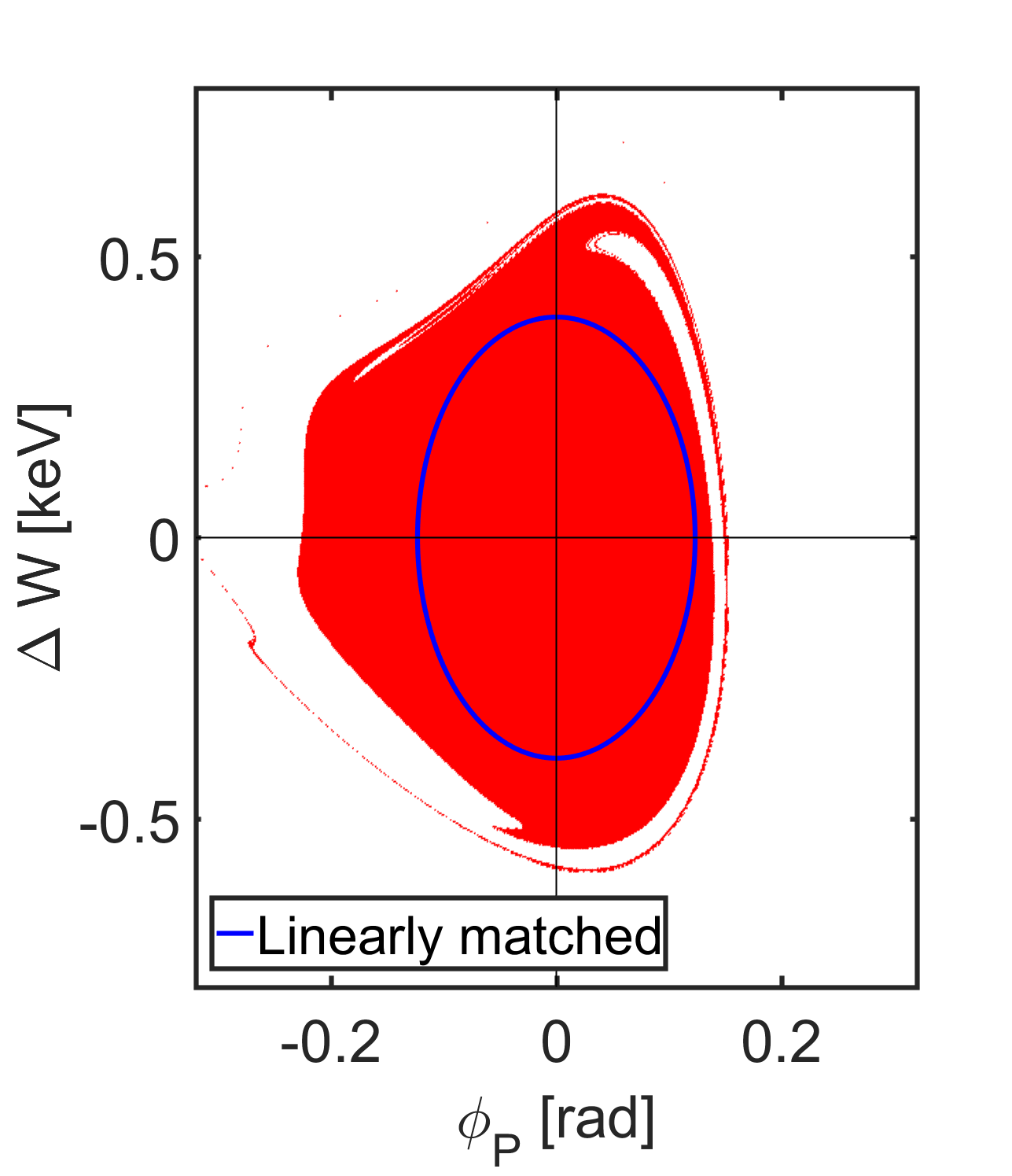}
\includegraphics[width=0.225\textwidth]{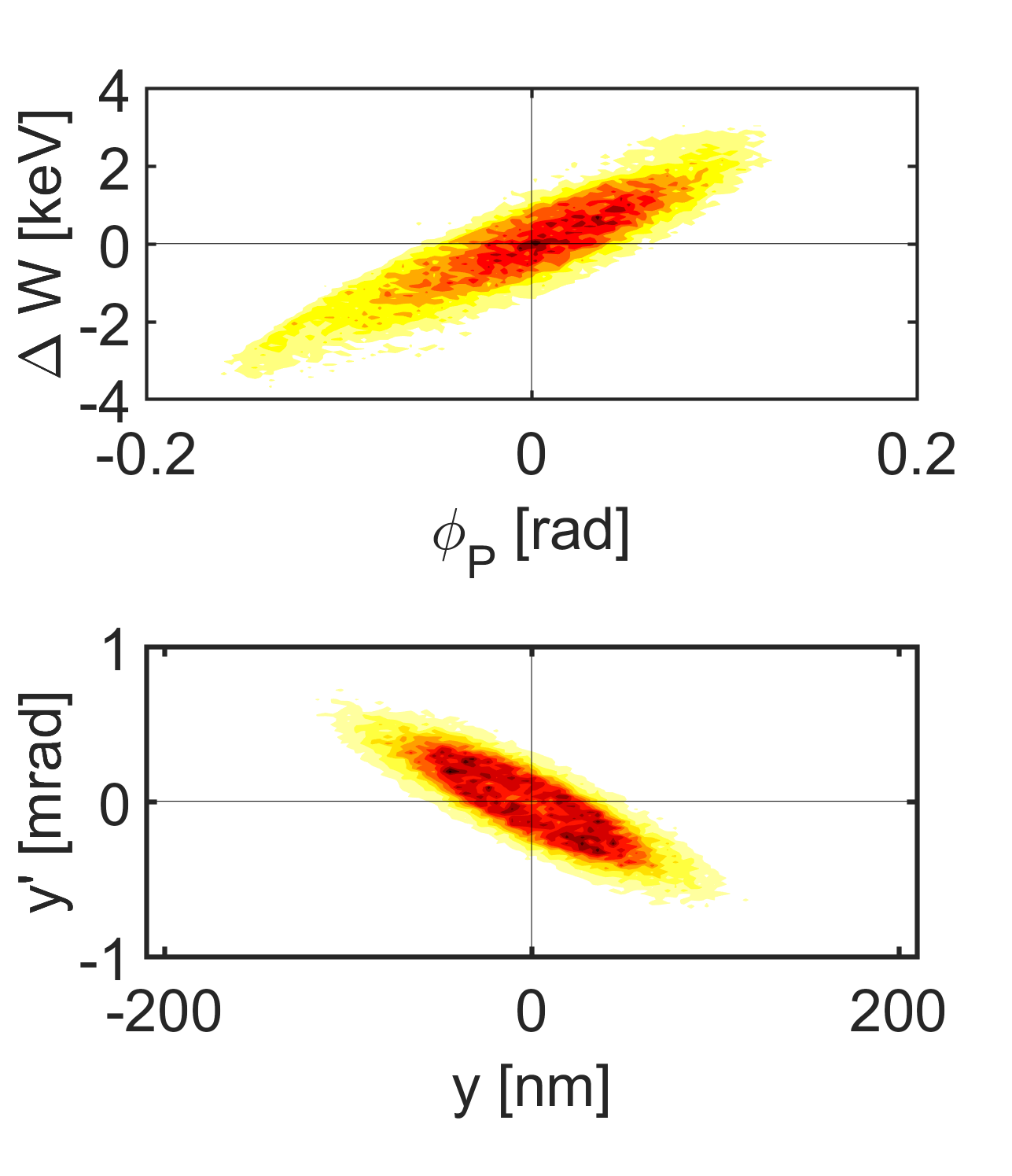}
\caption{Phase space after acceleration of a Gaussian bunch up to 1 MeV (right) and transmittable initial longitudinal distribution at 83 keV for $y=y'=0$ (left). 
The blue ellipse represents a linearly matched bunch with a total bunch length $4\sigma_z=40$ nm.}
\label{FinalPhaseSpace}
\end{figure}
The phase space density at top energy is plotted in Fig.~\ref{FinalPhaseSpace}, where $\Phi_P$ and $\Delta W$ are the longitudinal coordinates in the co-moving (Galilean) laboratory frame. 
As in Fig.~\ref{betafunctioncontour}, the initial particle positions in Fig.~\ref{FinalPhaseSpace} (left) are arranged 
on a cartesian grid, and only the ones that make it to 1~MeV are drawn in red. 
The blue ellipse corresponds to an initially matched bunch adjusted to the area of the surviving particles. 
Note that this size is slightly reduced at finite transverse emittance, thus we chose $\sigma_z=10$ nm. 
Below this bunch length, the transmission depends only on the transverse emittance. As the particle losses occur mostly in the beginning, the interaction length or the energy gain is scalable up to the available laser power.

\begin{figure}[t]
\centering
\includegraphics[width=0.42\textwidth]{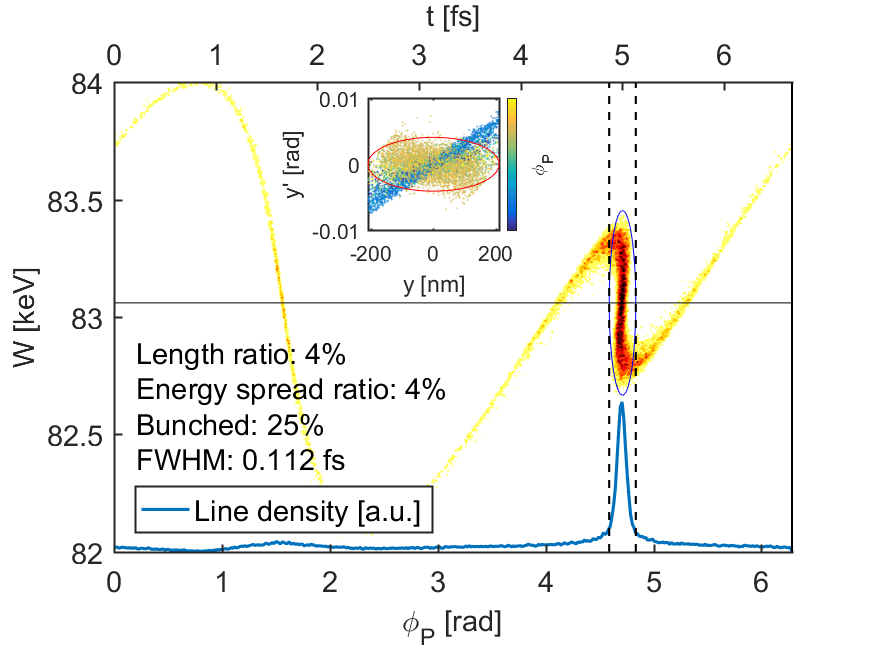}
\caption{Phase space after APF bunching. The initial beam parameters for the accelerator given by the ellipses are met.}
\label{Fig:Buncher}
\end{figure}

The APF scheme discussed here can also be used for bunching. Creating and removing sinusoidal energy spread (see~\cite{Niedermayer2018SupportingMaterial} and video therein) results in extremely short (attosecond) bunch lengths at acceptably low energy spread. 
The phase alternation additionally provides transverse confinement. 
The particles not captured are defocused, while the captured ones remain at small longitudinal and transverse amplitudes, within the limits of Liouville's theorem.
The phase space after the buncher is plotted in Fig.~\ref{Fig:Buncher}, where the ellipses are matched  for the accelerator in Fig.~\ref{envelope} (the blue ellipses in Figs.~\ref{Fig:Buncher} and~\ref{FinalPhaseSpace} are identical). The initial energy spread is $\sigma_{\Delta W}=16$~eV, where the initial longitudinal emittance equals the final one in the ellipse and $25\%$ is captured. The duration is decreased to $4\%$ $(\approx 260\,\mathrm{as})$ whereas the energy spread is increased by the same ratio. The initial CS-functions are determined by inverse mapping of the desired final values for the accelerator. Additionally to the injection into DLAs, these short bunches are also very promising for ultrafast time-resolved electron microscopy.              

In conclusion, we have developed a scheme that makes DLA fully scalable. The entire accelerator or parts, such as a single focusing stage or the buncher can now be experimentally approached.
Acceleration of electrons from available sources up to the MeV range with gradients of several 100 MeV/m  
works with transmission rates well above $90\%$. The admissible synchronous phase is determined by the available bunch length at injection. Our bunching scheme provides these attosecond bunches with the matched energy spread and reasonable capture rate of $25\%$.
In principle, fully adiabatic bunching as in the RFQ is also possible. This would, however, require a larger total length.
The APF scheme can also be scaled to higher energies, where smaller beam size and larger physical apertures due to longer roll-off of the evanescent acceleration fields will ease the requirements.

\begin{acknowledgments}
U. N. would like to thank Holger Podlech for discussions on APF.
This work is funded by the Gordon and Betty Moore Foundation (Grant GBMF4744 to Stanford) and the German Federal Ministry of Education and Research (Grant FKZ:05K16RDB).
\end{acknowledgments}

\bibliography{Mendeley}
\end{document}